# Probing charged impurities in suspended graphene using Raman spectroscopy

Zhen Hua Ni,[†] Ting Yu,[†] Zhi Qiang Luo,[†] Ying Ying Wang,[†] Lei Liu,[†] Choun Pei Wong,[†] Jianmin Miao,[‡] Wei Huang,[§] and Ze Xiang Shen[†*]

[†] *Division of Physics and Applied Physics, School of Physical and Mathematical Sciences, Nanyang Technological University, 21 Nanyang link, Singapore 637371*

[‡] *Micromachines Centre, School of Mechanical and Aerospace Engineering, Nanyang Technological University, 50 Nanyang Avenue, Singapore 639798*

[§] *Jiangsu Key Lab for Organic Electronics & Information Displays, Nanjing University of Posts and Telecommunications, 9 Wenyuan Road, Nanjing, China 210046*

**ABSTRACT**

Charged impurity (CI) scattering is one of the dominant factors that affect the carrier mobility in graphene. In this paper, we use Raman spectroscopy to probe the charged impurities in suspended graphene. We find that the 2D band intensity is very sensitive to the CI concentration in graphene, while the G band intensity is not affected. The intensity ratio between the 2D and G bands, $I_{2D}/I_G$, of suspended graphene is much stronger compared to that of non-suspended graphene, due to the extremely low CI concentration in the former. This finding is consistent with the ultra-high carrier mobility in suspended graphene observed in recent transport measurements. Our results also suggest that at low CI concentrations that are critical for device applications, the $I_{2D}/I_G$ ratio is a better criterion in selecting high quality single layer graphene samples than is the G band blue shift.

[*]Corresponding author: zexiang@ntu.edu.sg

The extremely high carrier mobility makes graphene a promising candidate for future electronic devices.[1] In practice, however, the carrier mobility of graphene varies from piece to piece [2,3,4] due to the different levels of charge impurity (CI) scattering present.[5,6] For example, the electron mobility of graphene can vary from $1\times10^3$ to $2\times10^4$ cm$^2$/Vs on SiO$_2$/Si substrate, which corresponds to a CI concentration range between $1.5\times10^{12}$ and $1\times10^{11}$ cm$^{-2}$.[4] It has been predicted that the carrier mobility of graphene can reach the ballistic limit of $\sim2\times10^6$ cm$^2$/Vs if the CI concentration can be decreased to $\sim10^{10}$ cm$^{-2}$.[5] In addition to the charged dopants from molecular adsorption and photoresist residues,[7] the substrate is another major source for charged impurities. Recent transport measurements on suspended graphene (SG) have revealed that the mobility of graphene can be dramatically enhanced to $\sim2\times10^5$ cm$^2$/Vs.[8,9] Such an enhancement is thought to be due to the absence of long range scattering from the random charged impurities in the substrate.[8] The experimental investigation of charged impurities in SG as well as comparison with those in non-suspended graphene (NSG) would therefore be desirable.

Raman spectroscopy has been widely applied in the study of graphene.[10-18] It can be used for determining graphene thickness,[10] monitoring dopant concentration,[11] measuring strain [12,13,14] and for probing the electronic structure of graphene and multilayer graphene.[15] In this study, we compare the Raman spectra of SG and NSG and find that the 2D band intensity of SG is much stronger. This is attributed to the extremely low CI concentration in SG ($<10^{11}$ cm$^{-2}$). A detailed study on many pieces of single layer graphene (SLG) suggests that at low CI concentrations that are critical for device applications, the intensity ratio between Raman 2D and G bands is a sensitive indicator of the level of charged impurities present.

**RESULTS AND DISCUSSION**

The process of fabrication of the SG samples is shown schematically in Figure 1. First, an $SiO_2$/Si substrate, which consists of a 285 nm-thick $SiO_2$ film on single crystal Si wafer, was spin-coated with ~10 μm thick photoresist (Figure 1a). Photolithography was then used to pattern holes into the photoresist (Figure 1b). After deep reactive-ion etching (DRIE) of the areas unprotected by the photoresist and subsequent removal of the photoresist, $SiO_2$/Si substrate with periodic structures were obtained (Figure 1c). The diameter of the holes (typically between 3 to 8 μm) depends on the original feature size on the photolithographic mask, while the depth of the holes depends on the duration of the DRIE. Graphene samples were prepared on the patterned substrates using the micromechanical cleavage technique (Figure 1d).[19] The probability of finding graphene sheets coving the holes is quite high because of the high concentration of holes. This makes the preparation of SG easy and efficient. As examples, Figures 1e, f and g show the typical optical images of three SG samples.

Figure 2a shows the optical image of a graphene sample on a patterned substrate, with a hole diameter of ~8 μm. The sample contains graphene sheets of different thicknesses. SLG was distinguished from 3-layer graphene from the width of the 2D Raman band. The former has a width of ~30 $cm^{-1}$ while the latter has a width of ~57 $cm^{-1}$, [10,19] which can also be seen from the Raman imaging constructed using the 2D bandwidth in Figure 2b. Part of the SLG is suspended over the hole while the remaining part is supported by the $SiO_2$/Si substrate. Hence our SG and NSG come from the same piece of SLG. Figures 2c and 2d show the Raman intensity mapping using the G and 2D bands, respectively. The dashed blue circles in the Raman imaging indicate the hole, i.e. the SG area. As we have shown, the Raman intensity for all the Raman bands from the NSG sample is enhanced as a result of the interference effect.[20] This explains the stronger G band Raman signal observed for the NSG sample (about twice the intensity compared to that of SG, as shown in Figure 2c). However, this is not the case for the 2D band intensity. The 2D Raman band intensity (Figure 2d) for the SG sample is stronger instead of weaker than that of the NSG sample. The difference is more clearly shown in the Raman image in Figure 2f, which is constructed using the $I_{2D}/I_G$ ratio. It can be seen that the $I_{2D}/I_G$ ratio varies significantly from 8.7 for SG to 3.9 for NSG. Figures 2g-2i show the Raman images of the $I_{2D}/I_G$ ratio of three more samples. Similarly, the $I_{2D}/I_G$ ratios of SG are much higher than those of NSG. We will explain this phenomenon later by considering the electron scattering in graphene. The samples used in this work were of high quality as indicated by the absence of an obvious disorder-induced D band in the Raman spectra of SG and NSG in Figure 2e.

It would be interesting to check whether there is any strain in SG.[14,21] To investigate this, the G band frequencies from different pieces of SG and NSG were recorded and the results are shown in Table 1. As we know, the frequency of the G band is very sensitive to strain. It red-shifts with a coefficient of 10 to 15 $cm^{-1}$/%strain due to the phonon deformation caused by the change in lattice constant.[14, 22] However, of the five SLG samples we studied, the G band frequencies of the SG are the same as those of the NSG, within an experimental error of ~1 cm$^{-1}$. The 2D band frequencies of the SG and NSG are also similar (results are not shown). This suggests that the strain in SG is negligible, which is consistent with the results of Berciaud et al.[23] Pereira et al.[21] suggested a method to open a transport bandgap in graphene by introducing local strain in it, which may be realized by placing graphene on local structures of substrates. From our results, it seems that noticeable strain (i.e. more than 1%) is not easily induced in graphene by simply placing it on local rough structures such as holes. This is reasonable as graphene is believed to be very stiff.[24, 25] One way to introduce a noticeable strain may be to anneal the SG sample, so that the graphene sheet can deform greatly at the edge of the holes. However, this is not within the scope of this work. In addition to the G band frequency, Table 1 also provides the 2D band width of SG and NSG. It can be seen that the 2D band of SG is much sharper than that of NSG. Such band narrowing is universal for all the samples we tested.

Next, we will focus on the abnormal change in the G and 2D band intensities of SG. The integrated intensity ratios of SG and NSG ($I_{SG}/I_{NSG}$) for different Raman bands are shown in Figure 3. The $I_{SG}/I_{NSG}$ of the G band centers at around 0.5, while that of 2D band has a much larger spread, which varies from 0.7 to 1.4. Our previous studies showed that the Raman intensity is strongly dependent on the interference of the laser and the Raman signals.[20] The Raman intensity of NSG (i.e. graphene on a 285 nm SiO$_2$ film on Si substrate) is greatly increased because of the substrate interference enhancement. The Raman intensity of the SG is also high because the optical constant $n$ ($n_{air}$=1) on both sides of graphene is smaller than that of graphene, which makes the interference and multiple reflections of the laser and Raman signal very efficient.[20] The calculated Raman intensity ratio between SG and NSG (on a 285 nm $SiO_2$/Si substrate) under 532 nm excitation is ~ 0.51, as indicated by the blue line in Figure 3. This value is very close to the $I_{SG}/I_{NSG}$ ratio of the G band. This suggests that the decrease in the G band intensity for SG is only due to different interference and multiple reflection conditions. On the other hand, the $I_{SG}/I_{NSG}$ ratio for the 2D band is much larger than the calculated value of 0.51. There must be factors in addition to interference and multiple reflections that contribute to such a discrepancy for the 2D band. Furthermore, such factors only affect the 2D band but not the G band.

The above phenomena can be understood by considering electron scattering in graphene.[5] The 2D band is a two-phonon Raman band which comes from the TO phonons around the K point of the Brillouin zone. It is active by the double resonance process which is described as follows:[26] 1) an excitation photon creates an electron-hole pair with similar energy at wave vector $k$. 2) electron-phonon scattering occurs with an exchanged momentum of $q$. 3) electron-phonon scattering takes place with an exchanged momentum $-q$, with reverse direction. 4) the electron-hole pair recombines. The matrix element of the process can be schematically represented as:[27]

$$M \sim \sum_{S0,S1,S2} \frac{\langle i|\hat{H}_{e-em}|s_0\rangle\langle s_0|\hat{H}_{e-ph}|s_1\rangle\langle s_1|\hat{H}_{e-ph}|s_2\rangle\langle s_2|\hat{H}_{e-em}|f\rangle}{(E_i - E_0 + 2i\gamma)(E_i - E_1 + 2i\gamma)(E_i - E_2 + 2i\gamma)} \quad (1)$$

where $|i\rangle$ and $|f\rangle$ are the initial and final states of the process, and $S_0$, $S_1$, $S_2$ are the intermediate states where an electron-hole pair is created. $E_i$ and $E_0...E_2$ are the energies of these states and $2\gamma$ is the inverse lifetime of the electron or hole due to collisions or scattering. $2\gamma$ is also known as the inelastic scattering rate. $\hat{H}_{e-em}$ and $\hat{H}_{e-ph}$ are the Hamiltonians describing the interaction of electrons with the electromagnetic field and with the phonons, respectively. The intensity of the 2D band can be expressed as:[28]

$$I_{2D} = \frac{(e^2/c)^2}{48\pi} \frac{v^2}{c^2} \frac{\omega_{in}^2}{\gamma^2} [\frac{9F_K^2}{M\omega_K v^2} \frac{\sqrt{27}a^2}{4}]^2 \quad (2)$$

Here, $v$ is the Fermi velocity, $a$ is the lattice constant of graphene, $M$ is the mass of the carbon atom, and $F_K$ is the coupling constant. $\omega_{in}$ and $\omega_K$ are the frequencies of the incident laser and the 2D phonon at around the K point, respectively. It is clear that $I_{2D}$ is proportional to $\frac{1}{\gamma^2}$, where $2\gamma$ is the electron or hole inelastic scattering rate as mentioned above. As the amount of charged impurities (i.e. the random charged impurities in the substrate) increases, the carrier density in graphene will also increase.[4, 6] Therefore, the probability of electron-electron collisions and the inelastic scattering rate $2\gamma$ also increases. According to equation (2), it is obvious that the 2D band intensity will decrease for NSG due to the charged impurities in the $SiO_2$ substrate. As a result, the $I_{SG}/I_{NSG}$ ratio of the 2D band will increase. Previous theoretical studies [5, 29] have revealed that CI scattering from the substrate is one of the major factors that changes the electron mobility of graphene. It has also been observed in transport measurements of SG that the mobility is greatly enhanced due to the absence of long range scattering of electrons or holes with substrate charged impurities.[8] Here, our Raman measurements on SG and NSG provide another evidence for the existence of substrate charged impurities.

On the other hand, the effect of substrate charged impurities on the G band intensity should be very weak. The G band originates from the $E_{2g}$ phonon, which has a wave vector of zero. Thus, the Raman process for the G band can be satisfied even under non-resonant conditions. As a result, the intensity of G band is expected to be insensitive to most of the external factors, such as polarization, carrier concentration and so on.[27] The effect of substrate charged impurities on the G band intensity hence can be ignored. This is consistent with our observation for SG and NSG. Accordingly, the intensity ratio of the 2D band to the G band, $I_{2D}/I_G$, would be a good indication of the amount of charge impurities in graphene. Previous studies on SLG samples have revealed an overall decrease of $I_{2D}/I_G$ when the amount of charged impurities in graphene increases.[7] This is further support for our argument. In previous studies, the blue shift in the G band was used as a direct indication of doping or the presence of charged impurities.[11, 30] However, we did not observe any obvious blue shift for the G band frequency on NSG with respect to that of SG (Table 1). This is because the NSG samples in Table 1 are only lightly doped, as indicated by their G band frequencies (~1580 cm$^{-1}$). The blue shift in the G band at such low CI concentrations (<10$^{12}$ cm$^{-2}$) is only ~1 cm$^{-1}$, according to the results of gated-tuned Raman spectroscopy of graphene.[11, 30] According to our results, the change in the $I_{2D}/I_G$ ratio is more sensitive to the presence of charged impurities than is the shift of the G band frequency when graphene is lightly doped. We therefore propose that the $I_{2D}/I_G$ intensity ratio is a more effective criterion for the selection of intrinsic SLG sample at low impurity concentration levels (<10$^{12}$ cm$^{-2}$) for device applications. The squares in Figure 4 show the $I_{2D}/I_G$ ratios of tens of SLG samples (SG and NSG) at different CI concentrations. The CI concentrations in NSG are estimated from the G band blue shift.[11] The CI concentration in SG is estimated to be 10$^{10}$-10$^{11}$ cm$^{-2}$.[4, 8, 29] The higher the $I_{2D}/I_G$ ratio, the lower the CI concentration in graphene. Moreover, the change in the $I_{2D}/I_G$ ratio is more sensitive at low concentration levels. For comparison, the relation between the G band blue shift and CI concentration is also presented in Figure 4. Such a relation is obtained from the results of Raman spectroscopy of graphene with carrier concentrations tuned by gate voltage.[11] It is obvious that at low impurity concentrations ($<10^{12}$ cm$^{-2}$), the blue shift in the G band is very small and is not easily distinguished considering the experimental error. Finally, care must be taken when directly comparing the $I_{2D}/I_G$ ratio obtained by different excitation lasers, because this value is also affected by the excitation energy.[31]

**CONCLUSION**

In summary, Raman spectroscopy and imaging were used to study SG and NSG samples. The G band intensity of SG is found to be weaker than that of NSG, due to the substrate interference effect. On the other hand, the 2D band intensity of SG is much stronger than that of NSG due to the absence of substrate charged impurities in SG. This finding is consistent with the ultra-high mobility in suspended graphene observed in recent transport measurements. Our results also suggest that at low CI concentrations ($<10^{12}$ cm$^{-2}$), the intensity of the 2D band (or $I_{2D}/I_G$) is more sensitive to the presence of charged impurities than is the blue shift of the G band.[11] We therefore propose that the $I_{2D}/I_G$ ratio can be used as a good criterion for selecting intrinsic single graphene samples for device application, where higher $I_{2D}/I_G$ indicates a lower CI concentration and hence a higher carrier mobility.

# EXPERIMENTAL AND CALCULATION SECTION

## Raman spectroscopy and imaging

Raman imaging /spectroscopy were carried out using a WITEC CRM200 Raman system with 532 nm (2.33 eV) excitation. The laser power at the sample was kept below 0.5 mW to avoid laser induced heating.[18, 32] A 100× objective lens with NA=0.95 was used in the Raman experiments, with a laser spot size ~500 nm for 532 nm excitation. For the Raman image, the sample was placed on an *x-y* piezostage and scanned under the illumination of laser. The Raman spectra from every spot on the sample were recorded. The stage movement and data acquisition were controlled using ScanCtrl Spectroscopy Plus software from WITec GmbH, Germany. Data analysis was done by using WITec Project software.[19]

## Raman intensity calculation

The Raman intensity of NSG considering the interference of laser light and Raman signal is calculated by the following formulae:[20]

$$I = \int_0^{d_1} |t \cdot \alpha|^2 \Delta y \tag{3}$$

$$t = \frac{t_1 \cdot e^{\beta \cdot y} \cdot e^{-i \cdot \frac{2\pi \tilde{n}_1 \cdot y}{\lambda}} + t_1 \cdot r' \cdot e^{\beta \cdot (2d_1 - y)} \cdot e^{-i \cdot \frac{2\pi \tilde{n}_1 \cdot (2d_1 - y)}{\lambda}}}{1 + r_1 r' \cdot e^{-2i \cdot fi_1} \cdot e^{2 \cdot \beta \cdot d_1}} \tag{4}$$

$$\alpha = \frac{(e^{\beta y} + r' e^{\beta(2d_1 - y)}) t_1'}{1 + r_1 r' e^{2\beta d_1}} \tag{5}$$

where $t$ is the total amplitude of the electric field at a certain depth $y$ and $\alpha$ is a factor considering the multi-reflection of scattered Raman light in graphene at the interface of graphene/air and graphene/(SiO$_2$ on Si). $\beta = \dfrac{-2\pi \cdot k_1}{\lambda}$, ($k_1$ =1.3 is the extinction coefficient of graphite and $\lambda$ is the excitation wavelength) is a measure of the absorption in the graphene layers. $t_1 = \dfrac{2\tilde{n}_0}{\tilde{n}_0 + \tilde{n}_1}$, $t_1' = \dfrac{1 - r_1^2}{t_1}$ are transmission coefficients at the interface of air/graphene and graphene/air. $r' = \dfrac{r_2 + r_3 \cdot e^{-2i \cdot fi_2}}{1 + r_2 \cdot r_3 \cdot e^{-2i \cdot fi_2}}$ is the effective reflection coefficient of graphene/(SiO$_2$ on Si) interface. $r_1 = \dfrac{\tilde{n}_0 - \tilde{n}_1}{\tilde{n}_0 + \tilde{n}_1}$, $r_2 = \dfrac{\tilde{n}_1 - \tilde{n}_2}{\tilde{n}_1 + \tilde{n}_2}$, $r_3 = \dfrac{\tilde{n}_2 - \tilde{n}_3}{\tilde{n}_2 - \tilde{n}_3}$ are reflection coefficients at the interface of air/graphene, graphene/SiO$_2$ and SiO$_2$/Si. $fi_{1,2} = \dfrac{2\pi \cdot \tilde{n}_{1,2} \cdot d_{1,2}}{\lambda}$ are the phase differences when light passes through graphene and SiO$_2$, respectively. $\tilde{n}_0 = 1$, $\tilde{n}_1 = 2.6-1.3i$, $\tilde{n}_2 = 1.46$, $\tilde{n}_3 = 4.15-0.044i$, are refractive indices of air, graphite, SiO$_2$, and Si at 532 nm, respectively.[33] $d_1$=0.335 nm is the thickness of single layer graphene, $d_2$=285 nm is the thickness of SiO$_2$ and the Si substrate is considered as semi-infinite.

The Raman intensity of SG is calculated by simply changing $\tilde{n}_2$ and $\tilde{n}_3$ to the refractive index of air $\tilde{n}_0 = 1$. The calculated Raman intensity ratio of SG and NSG, $I_{SG}/I_{NSG}$, is ~0.51.

Table 1 The G band frequency of SG and NSG from five different samples. The 2D band widths of SG and NSG are also presented.

| Samples | G frequency (cm$^{-1}$) | | 2D width (cm$^{-1}$) | |
|---|---|---|---|---|
| | **SG** | **NSG** | **SG** | **NSG** |
| **1** | **1578.7 $\pm$ 1.3** | **1577.6 $\pm$ 1.2** | **28.1 $\pm$ 1.6** | **31.7 $\pm$ 1.8** |
| **2** | **1579.6 $\pm$ 0.8** | **1580.2 $\pm$ 0.6** | **29.5 $\pm$ 1.1** | **35.4 $\pm$ 1.7** |
| **3** | **1580.9 $\pm$ 0.9** | **1581.1 $\pm$ 0.4** | **28.0 $\pm$ 0.7** | **31.8 $\pm$ 0.9** |
| **4** | **1579.2 $\pm$ 1.7** | **1580.8 $\pm$ 1.3** | **27.6 $\pm$ 2.0** | **29.5 $\pm$ 1.8** |
| **5** | **1582.6 $\pm$ 1.2** | **1581.4 $\pm$ 0.8** | **26.1 $\pm$ 1.3** | **31.6 $\pm$ 1.4** |

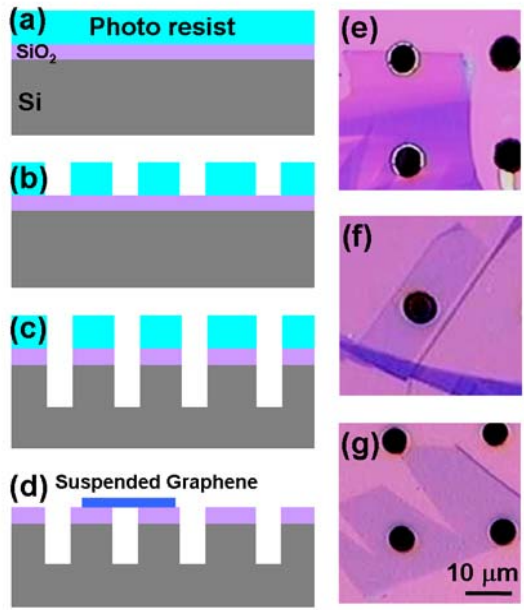

Figure 1. (a)-(d). Schematic diagrams for the preparation of suspended graphene. (a) A layer of photo-resist (10 μm thick) was deposited on the 285 nm $SiO_2$/Si substrate. (b) Photolithography was then used to pattern the photo-resist with 10 μm holes. (c) DRIE was used to etch the unprotected $SiO_2$ and Si. (d) Finally, suspended graphene was prepared on the patterned substrate. Figures (e) to (g) show three graphene samples with areas that are suspended.

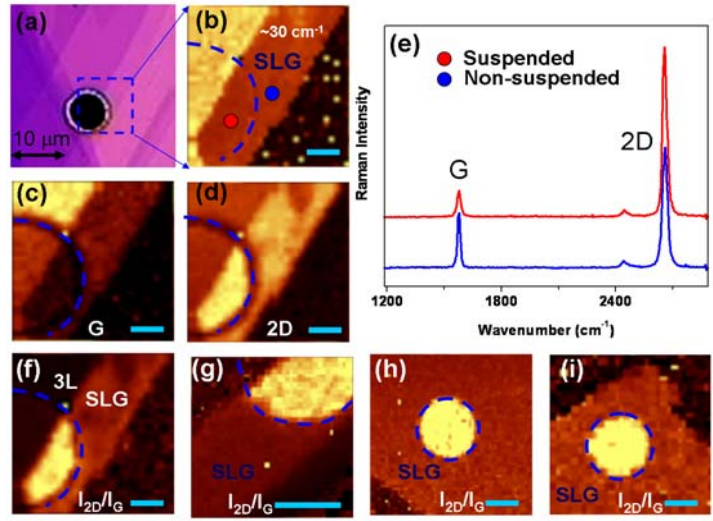

Figure 2. (a) Optical image of a graphene sheet on a patterned substrate covering a hole. (b) Raman imaging using the 2D band width. The dark strip with a 2D band width of ~30 cm$^{-1}$ is SLG. The bright area with 2D width of ~57 cm$^{-1}$ is three-layer graphene. (c) and (d) are the Raman imaging of G and 2D band intensity, respectively. (e) Raman spectra of SG and NSG taken from the red and blue dots in figure (d), respectively. (f) Raman imaging of the $I_{2D}/I_G$ ratio. (g)-(i) Raman images of $I_{2D}/I_G$ ratio of three more samples. The $I_{2D}/I_G$ ratios of SG are much higher than those of NSG. The scale bars in Raman images are 2 μm.

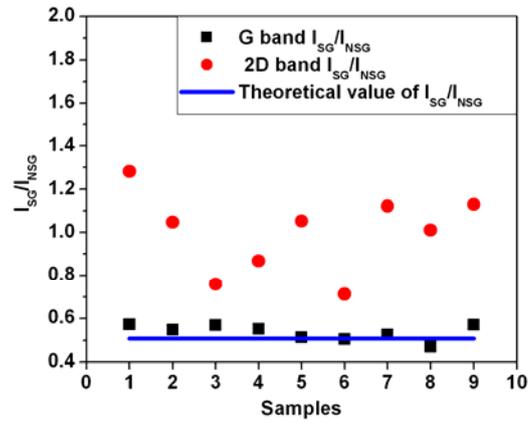

Figure 3. The G and 2D band integrated intensity ratio of suspended and non-suspended graphene. The blue line is the calculated value ( ~0.51) using the interference and multiple reflection model.[20] The results clearly indicate that while the G band intensity ratio $I_{SG}/I_{NSG}$ follow the calculated value well, the 2D band intensity ratio $I_{SG}/I_{NSG}$ does not.

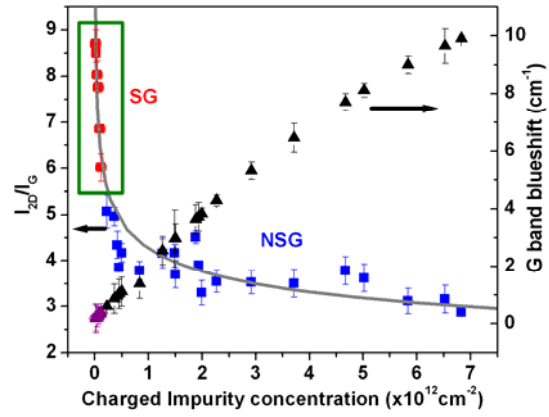

Figure 4. The G and 2D band integrated intensity ratio of SLG with different CI concentration: blue and red squares for NSG and SG, respectively. The solid line is a guide for the eye. For comparison, the relation between the G band blueshift and CI concentration is also presented: black and purple triangles for NSG and SG, respectively.